\documentclass[sigconf]{acmart}
\copyrightyear{2026}
\acmYear{2026}
\setcopyright{cc}
\setcctype{by}
\acmConference[SIGIR '26]{Proceedings of the 49th International ACM SIGIR Conference on Research and Development in Information Retrieval}{July 20--24, 2026}{Melbourne, VIC, Australia}
\acmBooktitle{Proceedings of the 49th International ACM SIGIR Conference on Research and Development in Information Retrieval (SIGIR '26), July 20--24, 2026, Melbourne, VIC, Australia}
\acmDOI{10.1145/3805712.3809943}
\acmISBN{979-8-4007-2599-9/2026/07}

\usepackage{tabularx} 
\usepackage{makecell}
\usepackage{threeparttable}
\usepackage{booktabs} 
\usepackage{multirow} 
\usepackage{array}    
\usepackage{graphicx} 
\usepackage{algpseudocode}
\usepackage{amsmath}
\usepackage{enumitem}
\usepackage{bbm}
\usepackage{balance}

\begin{document}

\title{L2Rec: Towards Dual-View Understanding of LLMs for Personalized Recommendation}

\author{Pingjun Pan}
\authornote{Both authors contributed equally to this research.}
\email{panpingjun@corp.netease.com}
\orcid{0009-0007-9892-9772}
\affiliation{%
  \institution{Netease Cloud Music}
  \city{Hangzhou}
  \country{China}
}

\author{Tingting Zhou}
\authornotemark[1]
\email{hzzhoutingting15@corp.netease.com}
\affiliation{%
  \institution{Netease Cloud Music}
  \city{Hangzhou}
  \country{China}
}

\author{Peiyao Lu}
\email{lupeiyao@corp.netease.com}
\affiliation{%
  \institution{Netease Cloud Music}
  \city{Hangzhou}
  \country{China}
}

\author{Tingting Fei}
\email{feitingting@corp.netease.com}
\affiliation{%
  \institution{Netease Cloud Music}
  \city{Hangzhou}
  \country{China}
}

\author{Hongxiang Chen}
\authornote{Hongxiang Chen is the corresponding author.}
\email{hzchenhongxiang@corp.netease.com}
\affiliation{%
  \institution{Netease Cloud Music}
  \city{Hangzhou}
  \country{China}
}

\author{Chuanjiang Luo}
\email{luochuanjiang03@corp.netease.com}
\affiliation{%
  \institution{Netease Cloud Music}
  \city{Hangzhou}
  \country{China}
}

\renewcommand{\shortauthors}{Pingjun Pan et al.}

\begin{abstract}
Adapting large language models (LLMs) for personalized recommendation requires aligning their general-purpose capabilities with user-specific preferences while effectively leveraging both behavioral and semantic signals. Existing approaches typically integrate these signals at either the input level (e.g., injecting behavioral embeddings into the token space) or the output level (e.g., contrastive alignment of separate encoders), suffering from distribution gaps or lack of end-to-end task supervision. In this work, we introduce \textbf{L2Rec}, which unifies behavioral and semantic understanding at the \emph{parameter level} of LLMs. Our key insight is that the same set of Transformer parameters can serve as a shared medium for both views: by applying view-specific, personalized low-rank perturbations via a Dual-view Personalized Mixture-of-Experts (DPMoE) mechanism, L2Rec enables a single LLM backbone to produce complementary behavioral and semantic adaptations for each user with minimal representation-level misalignment. An adaptive cross-view fusion module further integrates the dual-view outputs into a unified user preference. Experiments on four datasets show that L2Rec consistently outperforms state-of-the-art baselines, and online A/B testing on a large-scale industrial platform validates significant improvements in key engagement metrics.
\end{abstract}



\maketitle
\section{Introduction}

Large language models (LLMs) have emerged as a promising foundation for recommendation systems, owing to their rich world knowledge, contextual reasoning, and natural language understanding capabilities~\cite{ wu2024survey, lin2025can,zhang2025collm}. While traditional sequential methods~\cite{kang2018self, sun2019bert4rec, hou2022towards} effectively capture behavioral patterns through ID embeddings, they lack semantic understanding of item content. However, effectively deploying LLMs for personalized recommendation poses two intertwined challenges: (1) \emph{task alignment}---bridging the gap between the general-purpose language modeling objective and the user preference prediction goal of recommendation; and (2) \emph{signal integration}---jointly leveraging user behavioral patterns (e.g., click sequences) and semantic information (e.g., item descriptions), whose representations originate from different data sources.

Existing LLM-based recommendation methods can be broadly categorized by \emph{where} they integrate behavioral and semantic signals. The first category operates at the \textbf{output level}: LLMs serve as frozen feature extractors to generate semantic item embeddings, which are then consumed by separate downstream models~\cite{jia2025learn, ren2024representation, xi2024towards, wei2024llmrec}. While straightforward, this two-stage design decouples the LLM from recommendation supervision, preventing task-aligned gradient flow and creating an information bottleneck. The second category works at the \textbf{input level}: behavioral signals, typically pre-trained ID embeddings from collaborative filtering models, are projected into the LLM's token embedding space for joint processing~\cite{liao2024llara,zhang2025collm,zheng2024adapting}. Although this enables end-to-end training, it forces a single set of parameters to simultaneously handle heterogeneous representations—ID embeddings from sparse interaction matrices and token embeddings from dense text corpora—risking optimization conflicts as the model must reconcile gradients from two disparate sources. Critically, both paradigms  reconcile behavioral and semantic signals in the representation space, where distributional mismatches are hard to avoid.

We argue that a more effective integration point lies in the LLM's parameter space itself. Rather than aligning heterogeneous representations, we can adapt the same set of Transformer parameters to serve as a unified medium for both behavioral and semantic understanding. Building on this insight, we propose \textbf{L2Rec}, which applies view-specific, personalized low-rank adjustments to the LLM backbone via a Dual-view Personalized Mixture-of-Experts (DPMoE) mechanism. Concretely, DPMoE maintains a pool of LoRA-based experts~\cite{shazeer2017outrageously} whose activation is governed by a user-aware routing network that conditions on both user representations and contextual inputs. DPMoE instantiates two complementary adaptation pathways—SPMoE for semantic understanding and BPMoE for behavioral modeling—that share the same frozen backbone while producing view-specific adjustments tailored to each user. Following DeepSeekMoE~\cite{dai2024deepseekmoe}, shared experts capture cross-view common patterns while view-specific experts extract unique signals. An Adaptive Cross-view Fusion (ACF) module then dynamically integrates the dual-view outputs into a unified user preference. The entire framework is trained end-to-end with the LLM backbone frozen, updating only the lightweight DPMoE and ACF parameters.

Our main contributions are as follows:
\begin{itemize}[leftmargin=1em]
\item We identify \emph{parameter-level adaptation} as an effective alternative to input- or output-level fusion for integrating behavioral and semantic signals in LLM-based recommendation, avoiding the distributional misalignment inherent in representation-space approaches.
\item We propose L2Rec with a Dual-view Personalized Mixture-of-Experts  mechanism that applies user-aware, view-specific low-rank adjustments to a shared LLM backbone, enabling personalized dual-view adaptation through end-to-end parameter-efficient fine-tuning.
\item Extensive experiments on four benchmarks demonstrate consistent improvements over state-of-the-art baselines, and online A/B testing on a large-scale industrial platform confirms significant gains in user engagement metrics.
\end{itemize}

\section{Methodology}

\begin{figure}[t]
  \centering
  \includegraphics[width=1.0\linewidth]{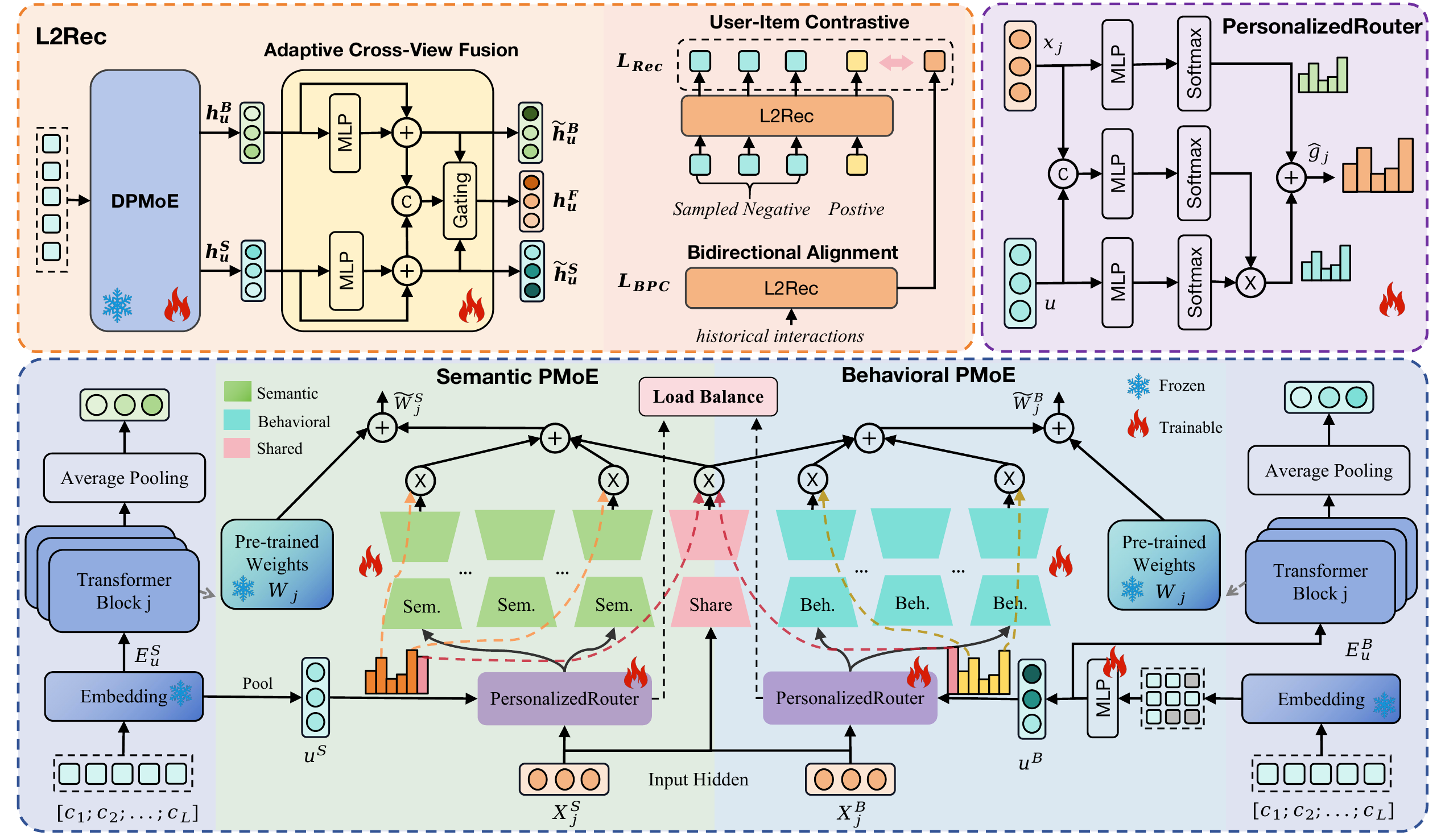}
  \caption{The overall framework of L2Rec.}
  \Description{L2Rec framework}
  \label{L2Rec framework}
\end{figure}

\subsection{Parameter-Level Dual-View Adaptation}

Existing methods integrate behavioral and semantic signals in the representation space, where distributional mismatches between the two views are difficult to reconcile. As illustrated in Figure~\ref{L2Rec framework}, L2Rec instead operates in the \emph{parameter space}: given a frozen LLM with parameters $\mathbf{W}$, we construct two parallel adaptation pathways by applying different low-rank adjustments $\Delta\mathbf{W}^S$ and $\Delta\mathbf{W}^B$ to the same backbone, yielding view-specific adapted parameters $\widetilde{\mathbf{W}}^S = \mathbf{W} + \Delta\mathbf{W}^S$ and $\widetilde{\mathbf{W}}^B = \mathbf{W} + \Delta\mathbf{W}^B$. Because both adjustments act on a shared parameter space grounded in the LLM's pre-trained knowledge, behavioral and semantic views share a common representational neighborhood, significantly reducing cross-view alignment difficulty.

\noindent\textbf{Dual-View Input Construction.}
Given a user's interaction sequence $\{(i_1, \mathbf{c}_1), \ldots, (i_L, \mathbf{c}_L)\}$, where $\mathbf{c}_k$ is the textual content of item $i_k$, we construct two complementary input views.
The \emph{semantic view} concatenates all item descriptions into a token sequence to capture fine-grained textual preferences:
\begin{equation}
\mathbf{E}_u^{S} = \text{Embed}([\mathbf{c}_1;\mathbf{c}_2; \ldots;\mathbf{c}_L]) \in \mathbb{R}^{T\times d}.
\end{equation}
The \emph{behavioral view} treats each item as an atomic unit to model sequential interaction patterns, analogous to SASRec~\cite{kang2018self}. Each item is compressed into a single vector via pooling over its token embeddings, and a lightweight projector $P_U$ maps it into a compatible space:
\begin{equation}
\mathbf{E}_{u}^B = P_{U}\big([\text{pool}(\text{Embed}(\mathbf{c}_i))]_{i=1}^{L}\big) \in \mathbb{R}^{L \times d}.
\end{equation}
The semantic view preserves token-level detail for understanding what items are about, while the behavioral view captures item-level sequential dynamics for modeling how user interests evolve---providing complementary signals at different granularities.

\subsection{Dual-View Personalized Mixture-of-Experts}

The key question is how to generate the adjustments $\Delta\mathbf{W}^S$ and $\Delta\mathbf{W}^B$ in a way that is both user-adaptive and parameter-efficient. A single LoRA adapter per view would apply a uniform adjustment to all users, ignoring the diversity of user preferences. We therefore propose DPMoE, which maintains a pool of LoRA experts and dynamically composes user-specific adjustments via personalized routing.

Behavioral and semantic preferences are not entirely disjoint---for instance, a user's affinity for ``sci-fi'' manifests both in click sequences and in item descriptions. We therefore partition the expert pool into shared experts $\mathcal{E}_{sh}$ and view-specific experts $\mathcal{E}_s$ (semantic), $\mathcal{E}_b$ (behavioral). Shared experts are always activated to provide a stable foundation of cross-view common knowledge, while view-specific experts are sparsely selected to capture user- and view-dependent specialization~\cite{dai2024deepseekmoe}. For a given adaptation pathway, the adjustment to the $j$-th parameter matrix $\mathbf{W}_j$ in a transformer block is:
\begin{equation}
\Delta\mathbf{W}_j = \sum_{i \in \mathcal{E}_{sh}} \mathbf{B}_{j,i} \mathbf{A}_{j,i} + \sum_{i \in \mathcal{E}_{v}} g_{j,i}\, \mathbf{B}_{j,i} \mathbf{A}_{j,i},
\label{eq:pmoe}
\end{equation}
where the first term aggregates shared experts that are always active, and the second term combines view-specific experts $\mathcal{E}_v \in \{\mathcal{E}_s, \mathcal{E}_b\}$ weighted by routing scores $g_{j,i}$. Each expert $i$ is parameterized by low-rank matrices $\mathbf{B}_{j,i} \in \mathbb{R}^{d \times r}$ and $\mathbf{A}_{j,i} \in \mathbb{R}^{r \times d}$ ($r \ll d$).

Standard MoE routers~\cite{shazeer2017outrageously} condition only on the current input $\mathbf{x}_j$, treating all users identically. In recommendation, however, the same input context should activate different experts for users with different preference profiles. We therefore design a three-signal router for the view-specific experts that combines: (i) \emph{context} $\mathbf{z}_j^c = \mathbf{x}_j$, capturing what is being processed; (ii) \emph{user} $\mathbf{z}_j^u = \mathbf{u}$, encoding who is being served; and (iii) \emph{interaction} $\mathbf{z}_j^f = [\mathbf{u};\mathbf{x}_j]$, modeling how user preferences interact with the current context. Each signal produces a score vector over $\mathcal{E}_v$ via a lightweight two-layer network:
\begin{equation}
\mathbf{g}_{j}^{(m)} = \mathrm{Softmax}\big(\mathbf{R}_{2,j}^{(m)} \cdot \mathrm{ReLU}(\mathbf{R}_{1,j}^{(m)} \mathbf{z}_{j}^{(m)} + \mathbf{b}_{1,j}^{(m)}) + \mathbf{b}_{2,j}^{(m)}\big),
\label{eq:router}
\end{equation}
where $m \in \{c, u, f\}$, and $\mathbf{R}_{1,j}^{(m)}$, $\mathbf{R}_{2,j}^{(m)}$ are learnable router weight matrices. The final routing weights use additive-multiplicative fusion with Top-$N$ sparsification:
\begin{equation}
\begin{split}
\hat{\mathbf{g}}_{j} &= \mathbf{g}_j^{c} + (\mathbf{g}_j^{u} + \mathbf{g}_j^{f}) \odot \mathbf{g}_j^{c}, \\
g_{j,i} &= \hat{g}_{j,i} \cdot \mathbbm{1}[\hat{g}_{j,i} \in \mathrm{Top\text{-}}N(\hat{\mathbf{g}}_{j})], \; i \in \mathcal{E}_v.
\end{split}
\label{eq:topn}
\end{equation}
where $\mathbbm{1}[\cdot]$ is the indicator function that retains only the Top-$N$ scoring experts and zeros out the rest.
This design ensures that context remains the primary routing signal, while user preferences act as a personalized corrective---modulating which experts are activated without overwhelming the context-dependent selection.

We denote Eqs.~\ref{eq:pmoe}--\ref{eq:topn} compactly as $\text{DPMoE}_j(\mathbf{u}, \mathbf{x}_j;\, \mathcal{E}_{sh},\, \mathcal{E}_v)$. The semantic and behavioral pathways are then instantiated as:
\begin{equation}
\begin{split}
\text{SPMoE}_j &= \text{DPMoE}_j(\mathbf{u}^S, \mathbf{x}_j^S;\, \mathcal{E}_{sh},\, \mathcal{E}_s), \\
\text{BPMoE}_j &= \text{DPMoE}_j(\mathbf{u}^B, \mathbf{x}_j^B;\, \mathcal{E}_{sh},\, \mathcal{E}_b).
\end{split}
\label{eq:dual_pmoe}
\end{equation}
where $\mathbf{u}^S{=}\text{pool}(\mathbf{E}_u^S)$ and $\mathbf{u}^B{=}\text{pool}(\mathbf{E}_u^B)$ are view-specific user representations. Through SPMoE and BPMoE, the same LLM backbone $\text{TRM}(\cdot)$ produces view-specific outputs:
$\mathbf{h}^S_u = \text{TRM}(\mathbf{E}_{u}^S;\, \text{SPMoE})$ and $\mathbf{h}^B_u = \text{TRM}(\mathbf{E}_{u}^B;\, \text{BPMoE})$.

\subsection{Training and Inference}

\noindent\textbf{Adaptive Cross-View Fusion (ACF).}
Since the dual-view outputs $\mathbf{h}^S_u$ and $\mathbf{h}^B_u$ already reside in a shared parameter space, lightweight fusion suffices. We apply residual projections followed by dynamic gating:
$\tilde{\mathbf{h}}^v_u = \mathbf{h}^v_u + P_v(\mathbf{h}^v_u)$ for $v \in \{B, S\}$, and
\begin{equation}
\delta = \sigma(W_g[\tilde{\mathbf{h}}^B_u;\tilde{\mathbf{h}}^S_u] + b_g), \quad
\mathbf{h}^{F}_u = \delta\, \tilde{\mathbf{h}}^B_u + (1{-}\delta)\, \tilde{\mathbf{h}}^S_u,
\end{equation}
where $\delta$ adaptively emphasizes each user's more informative view.

\noindent\textbf{Training Objectives.}
The model is trained with three complementary objectives. (1)~A \emph{contrastive recommendation loss} with in-batch negative augmentation:
\begin{equation}
\mathcal{L}_{\text{Rec}} = -\log 
\frac{e^{r_{u,i^{+}}}}
{e^{r_{u,i^{+}}} + \sum_{i^{-}\in\mathcal{N}} e^{r_{u,i^{-}}}},
\end{equation}
where $r_{u,i} = \text{sim}(\mathbf{h}^F_u, \mathbf{h}^F_i)/\tau$.
(2) Despite the shared backbone, input heterogeneity and expert specialization still induce representation drift. A \emph{Bidirectional Preference Contrastive (BPC) loss} that encourages consistent preference modeling across views by aligning $\tilde{\mathbf{h}}^B_u$ and $\tilde{\mathbf{h}}^S_u$ of the same user while separating different users:
\begin{equation}
\mathcal{L}_{\text{BPC}} =
-\log\frac{e^{r_{u}^{B \to S}}}{\sum_{u'} e^{r_{u,u'}^{B \to S}}}
-\log\frac{e^{r_{u}^{S \to B}}}{\sum_{u'} e^{r_{u,u'}^{S \to B}}}
+ \lambda \|\tilde{\mathbf{h}}^B_u - \tilde{\mathbf{h}}^S_u\|_2^2.
\end{equation}
(3)~A \emph{load-balancing loss} $\mathcal{L}_{\text{LB}}$ to prevent expert routing collapse. The overall objective is $\mathcal{L} = \mathcal{L}_{\text{Rec}} + \gamma\, \mathcal{L}_{\text{BPC}} + \beta\, \mathcal{L}_{\text{LB}}$.
During training, only the DPMoE experts and ACF parameters are updated while the LLM backbone remains frozen, ensuring parameter efficiency.

\noindent\textbf{Inference.}
The user representation $\mathbf{h}^F_u$ is obtained from ACF, and each candidate item is likewise encoded and fused via the same dual-view pipeline to obtain $\mathbf{h}^F_i$. The next item is predicted as $\hat{i}_u = \arg\max_{i \in \mathcal{I}}\, r_{u,i}$.

\section{Experiments}
\subsection{Experimental Settings}
\textbf{Dataset and Metrics.}  
We evaluate L2Rec on a large-scale \textbf{Industrial Dataset} and \textbf{Amazon Review} benchmark. 
The Industrial Dataset, sourced from a popular social platform, contains 42M chronological interactions (exposures, clicks, replies) from 1.5M users. Each item is associated with ten attributes, including demographic and content features. We utilize the first three months of logs for pre-training and the subsequent month for fine-tuning.
For Amazon Review, following RecFormer~\cite{li2023text}, we adopt a domain transfer setting: pre-training on seven source categories and fine-tuning on three target categories (\emph{Scientific}, \emph{Instruments}, \emph{Arts}). 
Evaluation follows the leave-one-out protocol~\cite{kang2018self}, reporting Recall@10 (R@10) and NDCG@10 (N@10). Table~\ref{tab:data_stats} details dataset statistics.

\noindent \textbf{Baselines.}  
We compare L2Rec with 7 competitive methods categorized into three groups:
(1) \textbf{ID-based}, including SASRec \cite{kang2018self} and BERT4Rec \cite{sun2019bert4rec};
(2) \textbf{Text-enhanced}, including S$^3$-Rec \cite{zhou2020s3}, UniSRec \cite{hou2022towards}, and RecFormer \cite{li2023text};
and (3) \textbf{LLM-based}, including LLaRA \cite{liao2024llara} ( Llama2-7B) and LEARN \cite{jia2025learn} (Baichuan2-7B).

\noindent \textbf{Implementation Details.}  
We employ Qwen3-0.6B~\cite{yang2025qwen3} as the backbone. L2Rec utilizes a dual-view PMoE with 9 experts (1 shared, 8 view-specific) implemented via LoRA ($r=8, \alpha=16$). The personalized router selects the Top-2 experts per view. Since the backbone is frozen, L2Rec updates only 32M parameters in DPMoE and ACF, roughly 5\% of the backbone size, supporting its parameter-efficient design. We optimize the model using AdamW with learning rates of 2e-4 (pre-training) and 1e-4 (fine-tuning), utilizing a contrastive loss with 10 randomly sampled negatives.
Maximum sequence lengths: $L=80, T=2048$ (Industrial) and $L=50, T=1024$ (Amazon). All hyperparameters are tuned on the validation set. Baselines use their original configurations and backbones; we additionally extend L2Rec to the same 7B backbones (Llama2-7B~\cite{touvron2023llama}, Baichuan2-7B~\cite{yang2023baichuan}) for  comparison. All experiments run on NVIDIA A100 GPUs.


\begin{table}[!t]
    \centering
    \caption{Overall statistics of the datasets.}
    \label{tab:data_stats}
    \renewcommand{\arraystretch}{1} 
    \resizebox{0.95\columnwidth}{!}{ 
    \begin{tabular}{lrrrrr}
        \toprule
        \textbf{Datasets} & \textbf{\#Users} & \textbf{\#Items} & \textbf{\#Inters.} & \textbf{Avg. n} & \textbf{Density }\\
        \midrule
        \textbf{Scientific} & 11,041 & 5,327 & 76,896 & 6.96 & $1.3\mathrm{e}{-3}$ \\
        \textbf{Instruments} & 27,530 & 10,611 & 231,312 & 8.40 & $7.9\mathrm{e}{-4}$ \\
        \textbf{Arts} & 56,210 & 22,855 & 492,492 & 8.76 & $3.8\mathrm{e}{-4}$ \\
           \textbf{Industrial } & 1,555,984 & 611,034 & 41,900,806 & 26.93 & $4.4\mathrm{e}{-5}$ \\
        \bottomrule
    \end{tabular}
    }
\end{table}

\begin{table}[!t]
    \centering
    \caption{Performance comparison across four datasets. Best results are in \textbf{bold}, second-best are \underline{underlined}.}
    \label{tab:performance_comparison}
    \resizebox{\columnwidth}{!}{
    \begin{tabular}{l|cc|cc|cc|cc}
        \toprule
        \multirow{2}{*}{Methods} & 
        \multicolumn{2}{c|}{Scientific} & 
        \multicolumn{2}{c|}{Instruments} & 
        \multicolumn{2}{c|}{Arts} & 
        \multicolumn{2}{c}{Industrial} \\
         \cmidrule(lr){2-3} \cmidrule(lr){4-5} \cmidrule(lr){6-7} \cmidrule(lr){8-9}
        & N@10 & R@10 & N@10 & R@10 & N@10 & R@10 & N@10 & R@10 \\
        \midrule
        SASRec \cite{kang2018self} & 0.0797 & 0.1305 & 0.0634 & 0.0995 & 0.0848 & 0.1342 & 0.1015 & 0.1167 \\
        BERT4Rec \cite{sun2019bert4rec} & 0.0790 & 0.1061 & 0.0707 & 0.0972 & 0.0942 & 0.1236 & 0.0816 & 0.0918 \\
        S$^3$-Rec \cite{zhou2020s3}  & 0.0451 & 0.0804 & 0.0797 & 0.1110 & 0.1026 & 0.1399 & 0.0874 & 0.1025 \\
        UniSRec \cite{hou2022towards} & 0.0862 & 0.1255 & 0.0785 & 0.1119 & 0.0894 & 0.1333 & 0.0881 & 0.1037 \\
        RecFormer \cite{li2023text} & 0.1027 & 0.1448 & 0.0830 & 0.1052 & \underline{0.1252} & 0.1614 & 0.1098 & 0.1183 \\
        LLaRA \cite{liao2024llara} & 0.1033 & 0.1502 & 0.0812 & 0.1078 & 0.1216 & 0.1630 & \underline{0.1133} & 0.1257 \\
        LEARN \cite{jia2025learn} & \underline{0.1060} & \underline{0.1594} & \underline{0.0878} & \underline{0.1240} & 0.1225 & \underline{0.1701} & 0.1127 & \underline{0.1299} \\
        \textbf{L2Rec} & \textbf{0.1145} & \textbf{0.1663} & \textbf{0.0912} & \textbf{0.1273} & \textbf{0.1318} & \textbf{0.1782} & \textbf{0.1198} & \textbf{0.1382} \\
        \midrule
        \%Improv. & +8.02\% & +4.33\% & +3.87\% & +2.66\%  & +5.27\% & +4.76\%  & +5.74\% & +6.39\% \\
        \bottomrule
    \end{tabular}
    }
\end{table}

\subsection{Overall Performance}
As shown in Table~\ref{tab:performance_comparison}, L2Rec consistently achieves the best performance across all four datasets, with \textbf{3.87\%--8.02\%} relative gains in N@10 over the strongest baseline. From the results, we observe a general progression across paradigms: text-enhanced methods tend to outperform pure ID-based methods, confirming the value of semantic information (e.g., on Arts, RecFormer achieves 0.1252 N@10, a +47.6\% gain over SASRec's 0.0848). LLM-based methods further improve but with diminishing gains (e.g., on Industrial, LEARN reaches 0.1127 N@10, improving only 2.6\% over RecFormer's 0.1098), suggesting that while LLMs provide a stronger foundation, input-level or output-level signal integration may not fully unlock their potential. L2Rec breaks through this ceiling via parameter-level adaptation, achieving consistent improvements across datasets of varying scales and domains.


\begin{table}[t]
\centering
\caption{Ablation Study on Components and Backbones}
\label{tab:ablation_study}
\resizebox{1.0\columnwidth}{!}{ 
\begin{tabular}{@{}l|cc|cc|cc@{}}
\toprule
\multirow{2}{*}{Variants} & \multicolumn{2}{c|}{Scientific} & \multicolumn{2}{c|}{Arts} & \multicolumn{2}{c}{Industrial} \\ 
\cmidrule(lr){2-3} \cmidrule(lr){4-5} \cmidrule(lr){6-7}
 & N@10 & R@10  & N@10 & R@10 & N@10 & R@10 \\ 
\midrule
\textbf{L2Rec (Qwen3-0.6B)} & \textbf{0.1145} & \textbf{0.1663}  & \textbf{0.1318} & \textbf{0.1782} & \textbf{0.1198} & \textbf{0.1382}\\ 
w/o Adapt & 0.0591 & 0.0868  & 0.0338 & 0.0600 & 0.0726 & 0.0925  \\
w/o Semantic & 0.1011 & 0.1456  & 0.1205 & 0.1599  & 0.0984 & 0.1126 \\
w/o Behavioral & 0.1062 & 0.1616 & 0.1238 & 0.1643 & 0.1055 & 0.1167 \\ 
w/o BPC & 0.1081 & 0.1637 & 0.1245 & 0.1665 & 0.1074 & 0.1206 \\ 
w/o PR & 0.1122 & 0.1649 & 0.1287 & 0.1755 & 0.1165 & 0.1361 \\ 
\midrule
L2Rec (Llama2-7B) & 0.1158 & 0.1668 & 0.1342 & 0.1802 & 0.1205 & 0.1391 \\
L2Rec (Baichuan2-7B) & 0.1159 & 0.1667 & 0.1344 & 0.1807 & 0.1213 & 0.1401 \\
\textbf{L2Rec (Qwen3-8B)} & \textbf{0.1178} & \textbf{0.1695} & \textbf{0.1362} & \textbf{0.1821} & \textbf{0.1232} & \textbf{0.1415} \\
\bottomrule
\end{tabular}
}
\end{table}

\begin{figure}[t]
  \centering
  \includegraphics[width=1.0\linewidth]{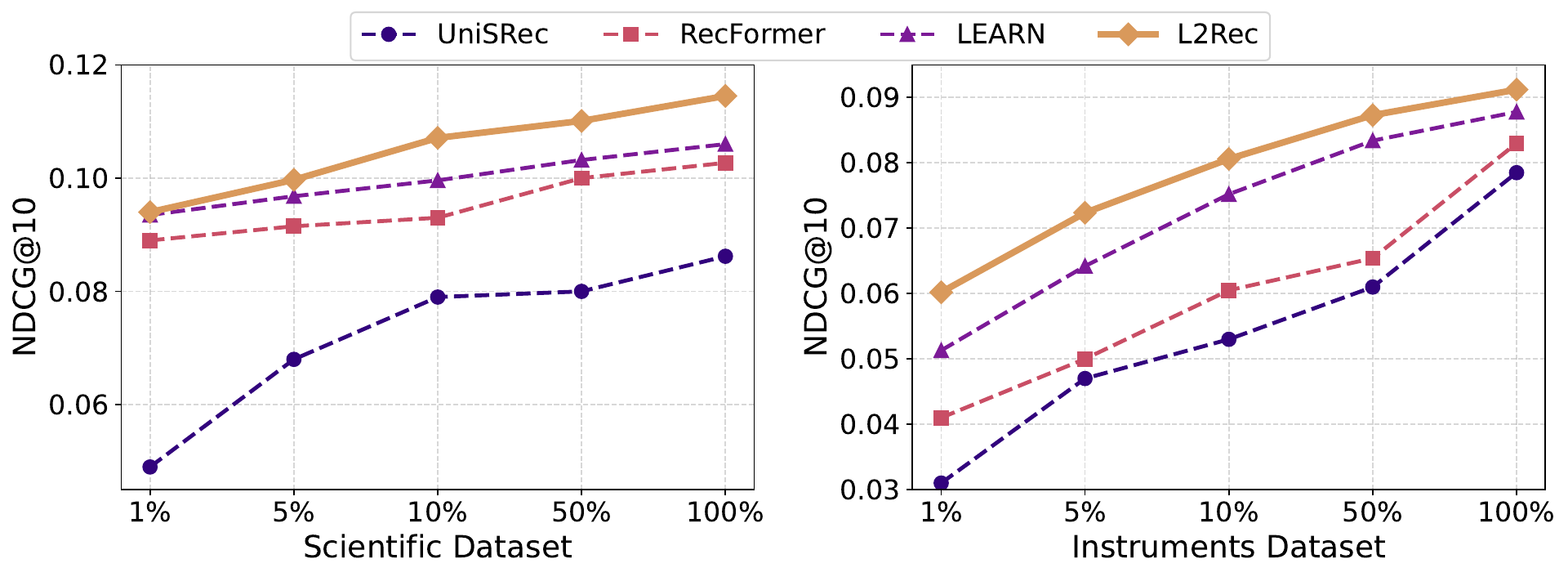}
  \caption{Data efficiency analysis under limited data.}
  \label{fig:low_resource}
\end{figure}

\begin{figure}[t]
\includegraphics[width=1.0\linewidth]{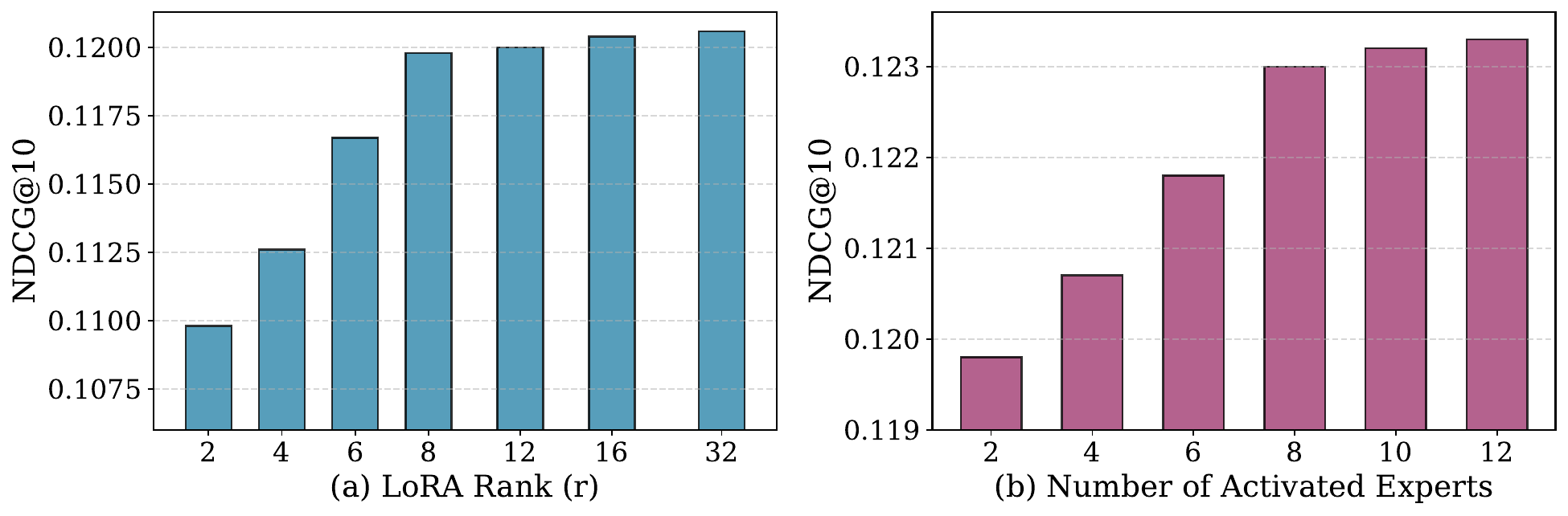}
  \caption{Hyperparameter Analysis of L2Rec}
  \label{fig:hyperparamaters}
\end{figure}

\subsection{Ablation Study}
Table~\ref{tab:ablation_study} presents ablation results on model components and backbone architectures across three datasets.

\noindent \textbf{Component Analysis.} 
\emph{w/o Adapt} removes all task-specific adaptation and reduces the model to the frozen Qwen3-0.6B backbone, causing severe degradation (e.g., N@10: 0.1198$\to$0.0726 on Industry, $-$39.4\%), which confirms that aligning LLMs with recommendation objectives is indispensable.
Within the full framework, \emph{w/o Semantic} decreases N@10 by 11.7\% (0.1145$\to$0.1011), 8.6\%, and 17.9\% on Scientific, Arts, and Industry respectively, while \emph{w/o Behavioral} leads to 7.2\%, 6.1\%, and 11.9\% drops, validating that both views provide complementary, non-substitutable signals.
\emph{w/o BPC} degrades N@10 by 5.6\%--10.4\%, confirming that cross-view preference consistency is critical for stabilizing dual-view adaptation.
\emph{w/o PR}, which replaces the personalized router with a user-agnostic router, yields 2.0\%--2.8\% N@10 declines, showing that user-aware expert activation enables more effective specialization.


\noindent \textbf{Impact of Backbone Architectures.}
We investigate whether L2Rec's gains stem from its DPMoE design or merely from backbone capacity. Since Table~\ref{tab:performance_comparison} follows each baseline's original backbone, the comparison is not strictly backbone-controlled, and the strong performance of Qwen3-0.6B may partially reflect differences in backbone quality and recency. To control for this factor, we instantiate L2Rec on the same larger backbones used by prior LLM-based recommenders. As shown in Table~\ref{tab:ablation_study}, scaling the backbone from Qwen3-0.6B to 8B yields consistent gains (e.g., N@10: 0.1198$\to$0.1232 on Industry). More importantly, under matched backbones, L2Rec(Llama2-7B) and L2Rec(Baichuan2-7B) consistently outperform LLaRA and LEARN by +6.4\%--12.1\% and +7.6\%--9.7\% in N@10, respectively, showing that the improvements mainly come from the proposed adaptation design rather than backbone choice alone.

\subsection{Performance under Limited Training Data}
\label{sec:low_resource}
We compare L2Rec with UniSRec, RecFormer (text-enhanced), and LEARN (LLM-based) on Scientific and Instruments across training ratios $\{1\%, 5\%, 10\%, 50\%, 100\%\}$. As shown in Figure~\ref{fig:low_resource}, L2Rec outperforms all baselines at every ratio. Text-enhanced methods degrade steeply under data scarcity, whereas both LLM-based methods exhibit flatter curves, confirming that the superior generalization  of LLMs provides strong initialization. The critical comparison is between L2Rec and LEARN, which isolates the effect of adaptation strategy given comparable LLM foundations. On Scientific, L2Rec with only 10\% data (0.1071) already surpasses LEARN at 100\% (0.1060); on Instruments, L2Rec outperforms LEARN by +17.3\% at 1\% (0.0602 vs.\ 0.0513). These results suggest that DPMoE's view-specific adaptation more effectively leverages the LLM backbone than LEARN's single-pathway design under limited supervision.

\subsection{Parameter Analysis}
We further examine key hyperparameters on the industrial dataset (Figure~\ref{fig:hyperparamaters}).
For the LoRA rank $r$ (2–32), performance improves as $r$ increases but gradually converges when $r$ becomes large, indicating that higher-rank adaptation enhances representation capacity with diminishing returns.
A similar trend is observed for the number of activated experts in each view: increasing the Top-N activated experts per view from 2 to 12 leads to sustained improvement with a gradually saturating effect. Overall, while higher ranks and increased experts yield steady gains, the diminishing returns observed imply that moderate settings may provide the best cost–benefit trade-off in practice.

\subsection{Online A/B Testing}
We conducted a one-month online A/B test on the homepage feed of a large-scale social platform ($\sim$1.5M DAU) from April to May 2025, with 6\% of user-level traffic allocated to the treatment group, while the remaining traffic was served by the control. The production baseline is a long-term optimized DLRM-based matching system. L2Rec serves as an upstream representation module: its dual-view user and item embeddings are fed into the downstream matching and ranking pipeline, with all other components kept unchanged, thereby isolating the effect of the proposed parameter-level dual-view adaptation. Compared with the baseline, L2Rec achieves \textbf{+9.24\%} in click-through rate and \textbf{+3.15\%} in reply rate, and both gains are statistically significant under a two-sample t-test on user-level metrics ($p < 0.01$), confirming that parameter-level dual-view adaptation produces higher-quality user--item representations that translate into significant engagement gains in production.

\section{Conclusion}

We present L2Rec, which tackles the intertwined challenges of task alignment and signal integration in LLM-based recommendation. Instead of reconciling behavioral and semantic signals in representation space as existing input- or output-level methods do, L2Rec operates in the parameter space: a frozen LLM backbone is equipped with view-specific adaptation pathways via DPMoE, where shared experts capture cross-view commonalities and view-specific experts specialize through personalized routing. Experiments on three public and one industrial dataset demonstrate consistent improvements over state-of-the-art baselines, and online A/B testing confirms substantial gains in key engagement metrics.

\bibliographystyle{ACM-Reference-Format}
\balance
\bibliography{sample-base}

\end{document}